\documentclass[graybox]{svmult}

\usepackage{mathptmx}       % selects Times Roman as basic font
\usepackage{helvet}         % selects Helvetica as sans-serif font
\usepackage{courier}        % selects Courier as typewriter font
\usepackage{type1cm}        % activate if the above 3 fonts are
\usepackage{makeidx}         % allows index generation
\usepackage{graphicx}        % standard LaTeX graphics tool
\usepackage{multicol}        % used for the two-column index
\usepackage[bottom]{footmisc}% places footnotes at page bottom

\makeindex             % used for the subject index
\begin{document}

\title*{Eclipse Mapping: \\ Astrotomography of Accretion Discs}
\author{Raymundo Baptista}
\institute{Raymundo Baptista \at Departamento de F\'{\i}sica, UFSC, Campus
  Trindade, 88040-900, Florian\'opolis, Brazil, \email{raybap@gmail.com}}
\maketitle

\abstract*{The Eclipse Mapping Method is an indirect
imaging\index{indirect imaging} technique that transforms the shape of the
eclipse \index{eclipse} light curve\index{light curve} into a map of the surface 
brightness distribution of the occulted regions. Three decades of application
of this technique to the investigation of the structure, the spectrum and the
time evolution of accretion discs\index{accretion disc} around white
dwarfs\index{white dwarf} in cataclysmic variables\index{cataclysmic variable}
have enriched our understanding of these accretion\index{accretion} devices with
a wealth of details such as (but not limited to) moving
heating/cooling\index{heating wave} waves during outbursts in dwarf
novae\index{dwarf nova outburst}, tidally-induced spiral shocks\index{spiral shock}
of emitting gas with sub-Keplerian velocities, elliptical\index{elliptical disc}
precessing discs\index{precessing disc} associated to superhumps\index{superhump},
and measurements of the radial run of the disc viscosity\index{disc viscosity}
through the mapping of the disc flickering\index{flickering} sources. This chapter
reviews the principles of the method, discusses its performance, limitations,
useful error propagation\index{error propagation} procedures, as well as highlights
a selection of applications aimed at showing the possible scientific problems
that have been and may be addresses with it.}

\abstract{The Eclipse Mapping Method is an indirect
imaging\index{indirect imaging} technique that transforms the shape of the
eclipse light curve\index{light curve} into a map of the surface
brightness distribution of the occulted regions. Three decades of application
of this technique to the investigation of the structure, the spectrum and the
time evolution of accretion discs\index{accretion disc} around white
dwarfs\index{white dwarf} in cataclysmic  variables\index{cataclysmic variable}
have enriched our understanding of these accretion devices with
a wealth of details such as (but not limited to) moving
heating/cooling waves\index{heating wave} during outbursts in dwarf
novae\index{dwarf nova outburst}, tidally-induced spiral shocks\index{spiral shock}
of emitting gas with sub-Keplerian velocities, elliptical\index{elliptical disc}
precessing discs\index{precessing disc} associated to superhumps\index{superhump},
and measurements of the radial run of the disc viscosity\index{disc viscosity}
through the mapping of the disc flickering\index{flickering} sources. This chapter
reviews the principles of the method, discusses its performance, limitations,
useful error propagation\index{error propagation} procedures, and highlights
a selection of applications aimed at showing the possible scientific problems
that have been and may be addresses with it.}

\section{Context and Motivations}
\label{sec:1}

In weakly-magnetic cataclysmic variables\index{cataclysmic variable} (CVs) a
late-type star overfills its Roche lobe\index{Roche lobe} and feeds a companion
white dwarf\index{white dwarf} (WD) via an accretion disc\index{accretion disc},
which usually dominates the ultraviolet and optical light of the system \cite{bible}.
The temperatures\index{temperature} in CV\index{cataclysmic variable} discs may
vary from 5000~K in the outer regions to over 50000~K close to the disc centre,
and the surface density\index{surface density} may vary by equally significant
amounts over the disc surface.  Therefore, the spectrum emitted by different
regions of the accretion disc\index{accretion disc} may be very distinct.
Furthermore, aside of the component stars and the accretion
disc\index{accretion disc}, additional sources contribute to the integrated
light from the binary: a bright spot\index{bright spot} (BS) or
stream\index{accretion stream} forms where the gas from the
mass-donor\index{donor star} star hits the outer edge of the disc,
tidally-induced spiral structures\index{spiral shock} might
develop in the outer regions of extended discs \cite{steeghs97,bhs00},
and a fraction of the transferred mass may also be ejected from the binary
in a wind\index{wind} from the disc surface \cite{b95,knigge97}.
Because what one directly observes is the combination of the spectra emitted
from these diverse regions and sources, the interpretation of 
CVs\index{cataclysmic variable} observations is usually plagued by the ambiguity
associated with composite spectra.  The most effective way to overcome these
difficulties is with spatially resolved\index{spatially resolved} data, in
which the light from the different sources might be disentangled.
However, at typical sizes of less than a solar radius and distances of hundreds
of parsecs, accretion discs\index{accretion disc} in
CVs\index{cataclysmic variable} are seen at angular diameters of tens
of micro-arcseconds, and spatially resolving them is well beyond the reach
of current direct imaging interferometric techniques, both at optical and
radio wavelength ranges. Thus, resolving an accretion disc\index{accretion disc}
in a CV\index{cataclysmic variable} is presently possible only via indirect
imaging\index{indirect imaging}.

Developed in the 1980's by Horne \cite{h85}, the Eclipse Mapping Method
is an indirect imaging\index{indirect imaging} technique which provides
spatially resolved\index{spatially resolved} observational constraints of
accretion discs\index{accretion disc} in CVs\index{cataclysmic variable} on
angular scales of micro-arcseconds. It assembles the information contained
in the shape of the eclipse \index{eclipse} into a map of the accretion
disc\index{accretion disc} surface brightness \index{surface brightness}
distribution. The following sections describe the technique and provide a
set of examples aiming at illustrating the wealth of possible applications.

\section{Principles \& Inner Workings}
\label{sec:2}

The three basic assumptions of the standard eclipse mapping\index{eclipse map}
method are:
(i) the surface of the secondary star is given by its Roche
equipotential\index{Roche potential},
(ii) the brightness distribution is constrained to the orbital
plane\index{orbital plane}, and
(iii) the emitted radiation is independent of the orbital phase. 
While assumption (i) is reasonably robust and always employed, the others
are simplifications that do not hold in all situations and may be relaxed
with three-dimensional (3D) eclipse mapping\index{eclipse map, 3D}
implementations (see Sect.\,\ref{3d-map}).

A grid of intensities centred on the WD, the eclipse map\index{eclipse map},
is defined in the orbital plane\index{orbital plane}.
One usually adopts the distance from the disc centre to the internal 
Lagrangian point\index{Lagrangian point}, $R_\mathrm{L1}$, as the length scale.
With this definition the primary lobe has about the same size and form for
any reasonable value of the binary mass ratio\index{mass ratio} $q$
(=$M_2/M_1$, where $M_2$ and $M_1$ are the masses of the
mass-donor\index{donor star} star and the WD, respectively) \cite{h85}.
If the eclipse map\index{eclipse map} is an $N$ points flat, square grid of
side $\lambda R_\mathrm{L1}$, each of its surface element (pixel) has an area 
$(\lambda R_\mathrm{L1})^2/N$ and an associated intensity $I_{j}$. 
The solid angle comprised by each pixel as seen from the earth is then
\begin{equation}
\theta^{2} = \left[ \frac{(\lambda R_\mathrm{L1})^2}{N} \frac{1}{d^2} \right] 
\cos\,i \; ,
\end{equation}
where $d$ is the distance to the system. The value of $\lambda$ defines 
the area of the eclipse map\index{eclipse map} while the choice of $N$
sets its spatial resolution\index{spatial resolution}.

The model eclipse light curve\index{light curve} $m(\phi)$ is derived from
the intensities in the eclipse map\index{eclipse map},
\begin{equation}
m(\phi) = \theta^{2}\, \sum_{j=1}^{\rm N} \, I_{j}V_j(\phi) \; .
\label{eq:2}
\end{equation}

The eclipse geometry $V_j(\phi)$ specifies the fractional visibility of
each pixel as a function of orbital phase and may include fore-shortening
and limb darkening \index{limb darkening} factors \cite{h93,r98,w94}.
The fractional visibility of a given pixel is obtained by dividing the
pixel into smaller tiles and evaluating the Roche potential
\index{Roche potential} along the line of sight for each tile to see if
the potential falls below the value of the equipotential that defines the
Roche surface\index{Roche potential} of the mass-donor\index{donor star} star.
If so, the tile is occulted. The fractional visibility of the pixel is then the
sum of the visible tiles divided by the number of tiles.

The eclipse geometry is determined by the inclination\index{inclination} $i$,
the binary mass
ratio $q$ and the phase of inferior conjunction $\phi_0$ \cite{h85,h93}.
These parameters set the shape and extension of the projected shadow of
the mass-donor\index{donor star} star in the orbital plane\index{orbital plane}.
As the binary phase $\phi$ changes during eclipse, \index{eclipse} the
resulting shadow rotates around the L1 point\index{Lagrangian point},
progressively covering/uncovering different parts of the accretion
disc\index{accretion disc}.
This creates a grid of criss-crossed ingress/egress arcs in the orbital
plane\index{orbital plane}. A pixel with coordinates $(x,y)$ within the region
covered by this grid disappears and reappears from eclipse \index{eclipse} at
a particular pair of binary phases $(\phi_i,\phi_e)$\footnote
 {Pixels outside the region covered by the grid of criss-crossed
  arcs are never eclipsed, and have no corresponding ingress/egress
  phases $(\phi_i,\phi_e)$. Accordingly, there is no information about
  the surface brightness\index{surface brightness} distribution of these
  unocculted regions in the shape of the eclipse. \index{eclipse}}.
Thus, the eclipse geometry sets the connection between the image space
(the eclipse map\index{eclipse map}) and the data space (the light
curve\index{light curve}).

\begin{figure}[t]
%\sidecaption
\includegraphics[bb=1.3cm 0.5cm 19.5cm 16cm,scale=.47,angle=270]
                  {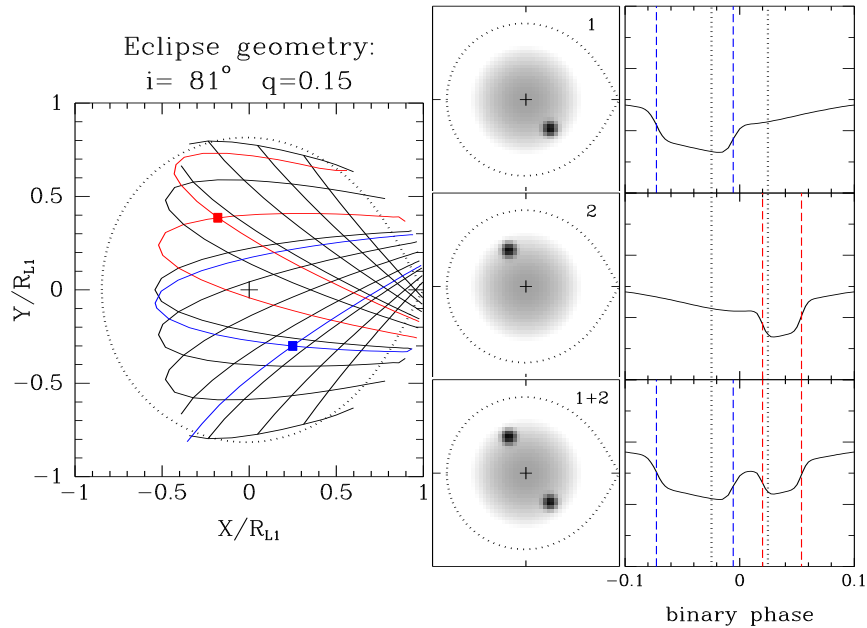}
\caption{Left-hand panel: A grid of criss-crossed ingress/egress arcs
  for the eclipse geometry $i=81^o$, $q=0.15$, and an eclipse
  map\index{eclipse map} with $\lambda=2$. The ingress/egress arcs
  for the blue and red pixels are shown in the respective colour.
  A dotted line depicts the primary Roche lobe\index{Roche lobe}; a cross
  marks the disc centre. The mass-donor\index{donor star} star is to the
  right of the panel and the stars rotate counter-clockwise (equivalently,
  the observer and the mass-donor\index{donor star}
  star shadow rotate clockwise). Middle-panels: grayscale brightness
  distributions with a faint, extended disc plus bright, narrow Gaussian
  spots at the positions of the blue and red pixels in the left panel.
  Right-hand panels: the model eclipse light curves\index{light curve}
  obtained by convolving the eclipse geometry in the left panel with the
  brightness distributions in the middle panels. Vertical dotted lines
  mark the ingress/egress of the disc centre; blue/red vertical dashed
  lines depict the ingress/egress phases of the blue/red pixels.
}
\label{fig:geom}
\end{figure}

\begin{figure}
\begin{center}
\includegraphics[bb=1cm 1cm 18cm 26cm,scale=.64]{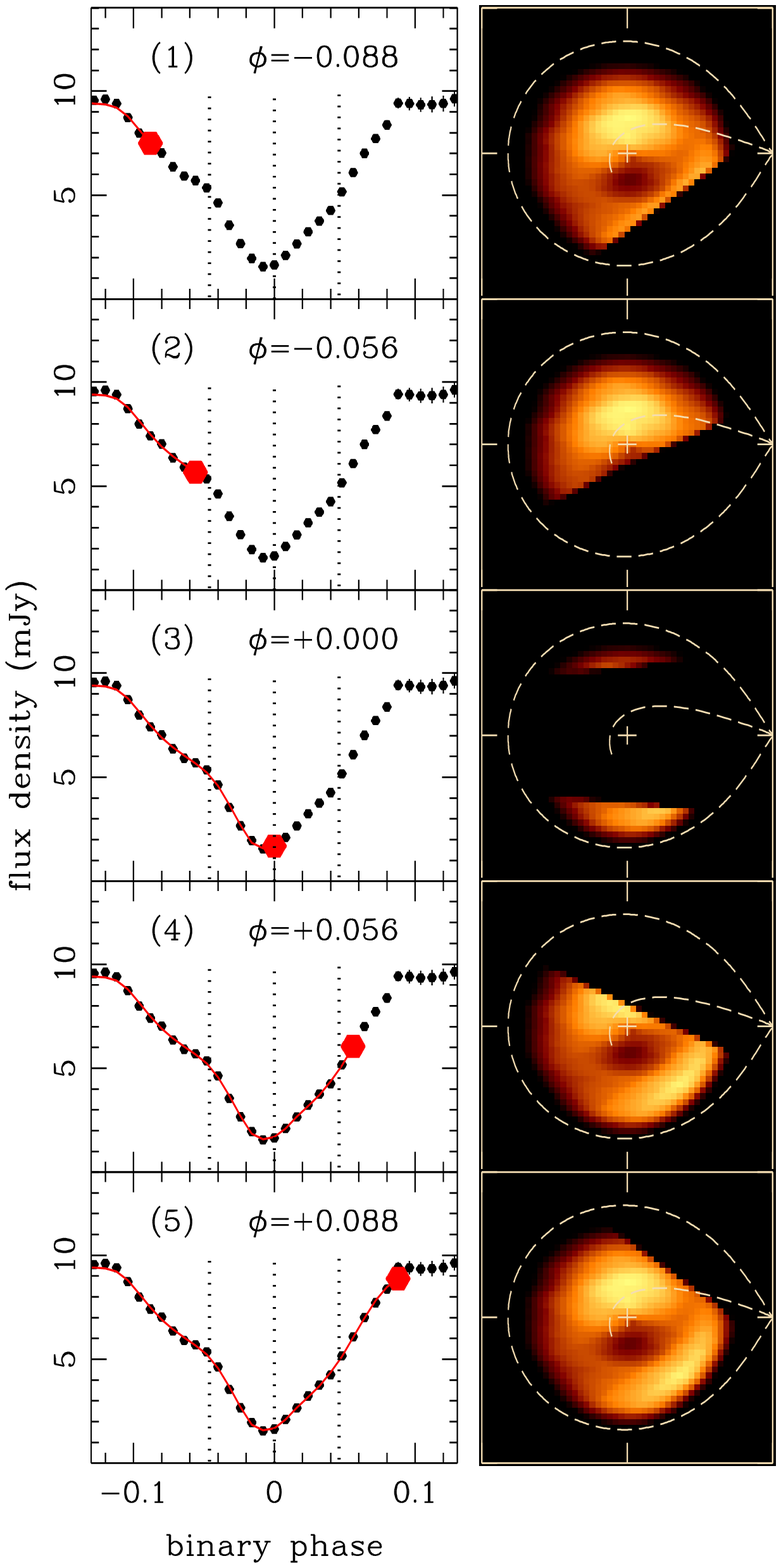}
\end{center}
\caption{ Simulation of a disc eclipse \index{eclipse} ($i=81^o$, $q=0.5$).
  Left-hand panels: data light curve (dots) and model light
  curve\index{light curve}
  (solid line) at five different binary phases (indicated in the upper 
  right corner). Vertical dotted lines mark mid-eclipse and the
  ingress/egress phases of the disc centre. Right-hand panels: visible
  parts of the best-fit eclipse map\index{eclipse map} (in a false colour
  blackbody logarithmic scale) at the corresponding binary phases.
  Dashed lines show the primary Roche lobe\index{Roche lobe} and the
  ballistic trajectory of the gas from the mass-donor\index{donor star}
  star; crosses mark the centre of the disc. The mass-donor\index{donor star}
  star is to the right of each panel; the stars
  and the accretion disc\index{accretion disc} gas rotate counter-clockwise.}
\label{fig:demo}
\end{figure}

Figure\,\ref{fig:geom} shows an example of eclipse geometry and
depicts the connection between the image and data spaces.
%
%%%%%%%%%% FIGURE 1 %%%%%%%%%%%%%
%
%%%%%%%%%%%%%%%%%%%%%%%%%%%%%%%%%
%
Information about the location of the occulted brightness sources is
embedded in the shape of the eclipse light curve\index{light curve}.
The width of the eclipse \index{eclipse} increases as the light source
moves closer to the L1 point\index{Lagrangian point}, whereas it is
displaced towards negative (positive) phases if the light source is in the
leading (trailing) disc side. For example, the phase width and range of
the eclipse \index{eclipse} in the upper right panel of Fig.\,\ref{fig:geom}
tells us that the corresponding light source (the blue pixel in the left-hand
panel of Fig.\,\ref{fig:geom}) is located in the leading, near side of the disc.
Figure~\ref{fig:demo} illustrates the simulation of the eclipse
\index{eclipse}  of a fitted brightness distribution while showing the
comparison between the resulting model light curve and the data light
curve\index{light curve}. The geometry in this case is $i=81^o$ and $q=0.5$. 
%
%%%%%%%%%%%%%%%%%%%%%%%%%%%  FIGURE 2  %%%%%%%%%%%%%%%%%%%%%%%%%%%%%%
%%%%%%%%%%%%%%%%%%%%%%%%%%%%%%%%%%%%%%%%%%%%%%%%%%%%%%%%%%%%%%%%%%%%
%
The left-hand panels show the data light curve\index{light curve} (small
dots) and the model
light curve (solid line) as it is being drawn at five different orbital
phases along the eclipse \index{eclipse} (indicated in the upper right corner).
The right-hand panels show the best-fit disc brightness distribution and
how it is progressively covered by the dark shadow of the
mass-donor\index{donor star} star during the eclipse.
The slope of the light curve\index{light curve} steepens whenever a bright
source is occulted or reappears from eclipse. \index{eclipse}
The two asymmetric arcs of the brightness
distribution of this example are occulted at distinct phases, leading
to a V-shaped eclipse \index{eclipse} with two bulges at ingress and at egress. 
The flux at phase $\phi=0$ does not go to zero because a sizeable 
fraction of the disc remains visible at mid-eclipse.

\begin{figure}
%\sidecaption
\includegraphics[scale=.6]{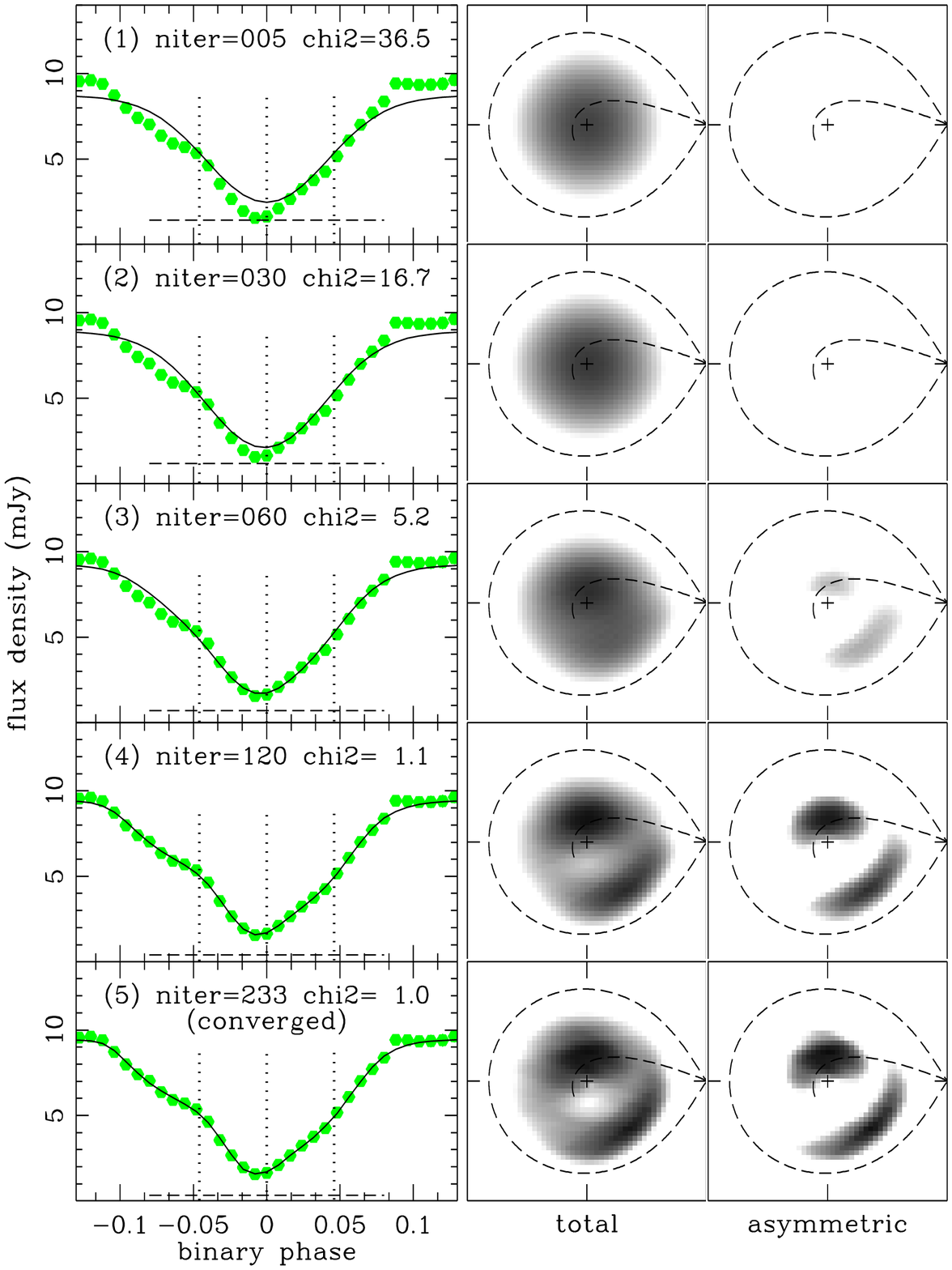}
\caption{Example of the iterative fit of a data light
  curve\index{light curve} with a MEM\index{maximum entropy}
  eclipse mapping\index{eclipse map} code \cite{bs93}. Left-hand panels:
  data (points) and
  model (solid line) light curves at five different stages of the iteration
  process towards convergence. Vertical dotted lines mark mid-eclipse and
  the ingress/egress of the disc centre. An horizontal dashed line depicts
  the uneclipsed component of the total flux. {\em niter} and {\em chi2}
  give the current iteration number and $\chi^2$ value, respectively.
  Middle panels: corresponding eclipse maps\index{eclipse map} in a
  logarithmic grayscale;
  darker regions are brighter. The notation is the similar to that of
  Fig.\,\ref{fig:demo}. Right-hand panels: the asymmetric component of
  the eclipse maps\index{eclipse map} in the middle panels.}
\label{fig:3}
\end{figure}

Given the eclipse geometry, a model light curve\index{light curve} can be
calculated for any assumed brightness distribution in the eclipse
map\index{eclipse map}. A computer code then iteratively adjusts the
intensities in the map (treated as independent parameters) to find the
brightness distribution the model light curve\index{light curve}
of which fits the data eclipse light curve\index{light curve} within
the uncertainties (Fig.\,\ref{fig:3}).
Because the one-dimensional data light curve cannot fully constrain a
two-dimensional map, additional freedom remains to optimise some map
property. A maximum entropy\index{maximum entropy}
(MEM\index{maximum entropy}) procedure \cite{sb84,s87} is used to select,
among all possible solutions, the one that maximises the
entropy\index{maximum entropy} of the eclipse map\index{eclipse map}
with respect to a smooth default map.

The entropy of the eclipse map\index{eclipse map} $p$ with respect to
the default map $q$ is defined as
\begin{equation}
S = - \sum_{j=1}^{N} \, p_j\,\ln \left( \frac{p_{j}}{q_{j}} \right) \; ,
\label{eq:1}
\end{equation}\index{entropy}
where $p$ and $q$ are written as
\begin{equation}
p_{j} = \frac{I_{j}}{\sum_{k}I_{k}}
\;\;\;\;\; , \;\;\;\;\; q_{j} = \frac{D_{j}}{\sum_{k}D_{k}} \, .
\end{equation}
Other functional forms for the entropy appear in the literature
\index{entropy} \cite{h93,bible}. These are equivalent to (\ref{eq:1})
when $p$ and $q$ are written in terms of proportions.

The default map $D_j$ is generally defined as a weighted average of the
intensities in the eclipse map\index{eclipse map},
\begin{equation}
D_{j} = \frac{\sum_{k} \, \omega_{jk}I_{k}}{\sum_{k} \, \omega_{jk}}\; ,
\end{equation}
where the weight function\index{weight function} $\omega_{jk}$ is
specified by the user.
A priori information about the disc (e.g., axi-symmetry) is included
in the default map via $\omega_{jk}$.
Prescriptions for the weight function $\omega_{jk}$ and their effect
in the reconstructions are discussed in Sect.\,\ref{sec:3}.
In the absence of any constraints on $I_j$, the entropy has a
maximum\index{maximum entropy} $S_{max}=0$ when $p_j=q_j$, or when
the eclipse map\index{eclipse map} and the default map are identical.

The consistency of an eclipse map\index{eclipse map} may be checked
using the $\chi^2$ as a constraint function,
\begin{equation}
\chi^{2} = \frac{1}{M}\:\sum_{\phi=1}^{M} \, 
\left( \frac{m(\phi)-d(\phi)}{\sigma(\phi)} \right)^2 =
\frac{1}{M}\:\sum_{\phi=1}^{M} \, r({\phi})^{2}\; ,
\label{eq:chi}
\end{equation}
where $d(\phi)$ is the data light curve\index{light curve},
$\sigma(\phi)$ are the corresponding uncertainties, $r(\phi)$ is the
residual at the orbital phase $\phi$, and $M$ is the number of data
points in the light curve\index{light curve}.
Alternatively, the constraint function may be a combination of the
$\chi^{2}$ and the R-statistics \cite{bs93},
\begin{equation}
R = \frac{1}{\sqrt{M-1}}\:\sum_{\phi=1}^{M-1}\,r({\phi}) \;
 r({\phi+1})\; ,
\label{eq:r}
\end{equation}
to minimise the presence of correlated residuals in the model light 
curve \cite{bs91}. For the case of uncorrelated normally distributed
residuals, the R-statistics has a Gaussian probability distribution 
function with average zero and unity standard deviation. Requiring the 
code to achieve an $R=0$, is equivalent to asking for a solution with 
uncorrelated residuals in the model light curve\index{light curve}.

The final MEM\index{maximum entropy} solution is the eclipse
map\index{eclipse map} that is as close as possible
to its default map as allowed by the constraint imposed by the light
curve and its associated uncertainties \cite{h93,r98}. In mathematical
terms, the problem is one of constrained maximisation, where the function
to maximise is the entropy\index{entropy} and the constraint is a
consistency statistics that measures the quality of the fitted model to
the data light curve\index{light curve}.
Different codes exist to solve this problem. Many of the eclipse
mapping\index{eclipse map} codes are based on the commercial
optimisation package MEMSYS \cite{sb84}.\index{MEMSYS}
Alternative implementations using conjugate-gradients algorithms
\cite{bs91,bs93}, CLEAN-like\index{CLEAN} algorithms \cite{spruit94} and
genetic algorithms\index{genetic algorithm} \cite{bob00} are also used.

Rutten et al. \cite{rpt92} found that the entropy\index{entropy}
function can be a useful tool to signal and to isolate the fraction
of the total light which is not coming from the accretion
disc\index{accretion disc} plane\index{orbital plane}.  They noted that
when the light curve\index{light curve} is contaminated by the presence
of additional light (e.g., from the mass-donor\index{donor star} star)
the reconstructed map shows a spurious structure in its outermost region.
This is because the eclipse mapping\index{eclipse map} method assumes
that all the light is coming from the accretion disc\index{accretion disc},
in which case the eclipse depth and width are correlated in the sense
that a steeper shape corresponds to a deeper eclipse. \index{eclipse}
The addition of an uneclipsed component in the light
curve\index{light curve} (i.e., light from a source other than the
accretion disc\index{accretion disc}) ruins this correlation.  To account
for the extra amount of light at mid-eclipse and to preserve the brightness
distribution derived from the eclipse \index{eclipse} shape the
algorithm inserts the additional light in the region of the map
which is least affected by the eclipse. \index{eclipse}
Since the entropy\index{entropy} measures the amount of structure in the
map, the presence of these spurious structures is flagged with lower
entropy values.  The correct offset level may be found by comparing a
set of maps obtained with different offsets and selecting the one with
highest entropy\index{maximum entropy}.  Alternatively, the value of
the zero-intensity level can be included in the mapping algorithm
as an additional free parameter to be fitted along with the intensity
map in the search for the MEM\index{maximum entropy} solution
\cite{b95,r94}. A detailed discussion on the reliability and
consistency of the estimation of the uneclipsed component can be
found in \cite{bsh}.

Figure\,\ref{fig:3} gives an example of the convergence process of a
MEM\index{maximum entropy} reconstruction. The code starts from a
flat map and quickly evolves towards an axi-symmetric Gaussian map
which reproduces the gross features of the data light
curve\index{light curve}. However, at this point the model light curve
is not a good match to the data ($\chi^2= 16.7$), failing to reproduce
the double stepped ingress/egress bulges of the eclipse shape.
\index{eclipse}  Several more iterations are required in order to match
these features in the data light curve\index{light curve},
which demands building two asymmetric bright arcs in the leading and
trailing sides of the disc. The reconstruction converges for a final unity
$\chi^2$ value after 233 iterations.
%
%%%%%%%%%% FIGURE 3 %%%%%%%%%%%%%
%
%%%%%%%%%%%%%%%%%%%%%%%%%%%%%%%%%

\section{Performance \& Limitations}
\label{sec:3}

\begin{figure}
%\sidecaption
\includegraphics[scale=.45,angle=270]{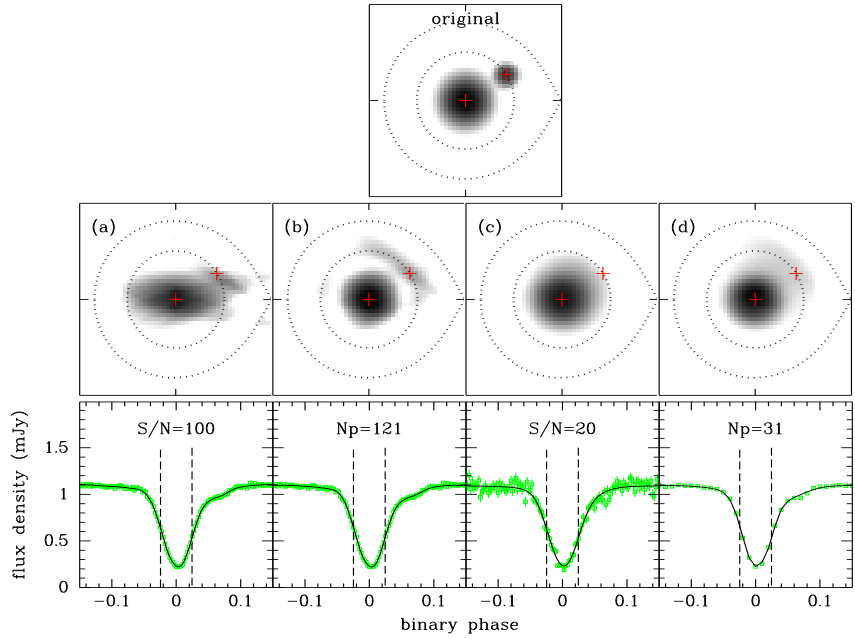}
\caption{Effects of the default map, phase resolution\index{phase resolution}
  (= number of data points in the light curve\index{light curve}) and
  signal-to-noise ratio (S/N) of the input data on the eclipse
  mapping\index{eclipse map} reconstruction. The top panel shows an
  artificial, test brightness distribution in a logarithmic grayscale.
  The notation is similar to that of Fig.\,\ref{fig:geom}. The middle panels
  show reconstructions of the test brightness distribution for different
  combination of parameters: (a) uniform default map on a light
  curve\index{light curve} with $N=121$ data points and $S/N=100$;
  (b) default of limited azimuthal smearing\index{azimuthal smearing}
  (Eq.\,\ref{eq:angle}), $Np=121$ and $S/N=100$; (c) default of
  Eq.\,\ref{eq:angle}, $Np=121$ and $S/N=20$; and (d) default of
  Eq.\,\ref{eq:angle}, $Np=31$ and $S/N=100$. The lower panels show
  the corresponding data (points) and model (solid line) light
  curves\index{light curve}.
  Vertical dashed lines mark the ingress/egress phases of the disc centre.}
\label{fig:4}
\end{figure}

A crucial aspects of eclipse mapping\index{eclipse map} is the selection
of the weight function\index{weight function} for the default map,
$\omega_{jk}$, which allows the investigator to steer the
MEM\index{maximum entropy} solution towards a determined type of disc map.
The choice $\omega_{jk}=1$ results in a uniform default map which leads
to the {\em most uniform eclipse map\index{eclipse map}} consistent with
the data. This happens not to be a good choice for eclipse
mapping\index{eclipse map}, because it
results in a map severely distorted by criss-crossed artefacts 
\cite{bob97,h85,spruit94} as the flux of point sources is spread along
their ingress and egress arcs (Fig.\,\ref{fig:4}(a)).
This led to the adoption of a weight function\index{weight function}
which sets the default map as an axi-symmetric average of the eclipse
map\index{eclipse map}, thereby yielding the
{\em most nearly axi-symmetric map} that fits the data \cite{h85}.
It suppresses the azimuthal information in the default map while keeping
the radial structure of $I_{j}$ on scales greater than a radial blur width
$\Delta_R$. This seems a reasonable choice for accretion
disc\index{accretion disc} mapping
because one expects the disc material to be roughly in Keplerian orbits,
so that local departures from axi-symmetry will tend to be smeared away by
the strong shear. This is a commonly used option and is also known as
the default map of full azimuthal smearing\index{azimuthal smearing}.

The full azimuthal smearing\index{azimuthal smearing} default results
in rather distorted reproduction of asymmetric structures such as a
bright spot\index{bright spot} at the disc rim\index{disc rim}.
In this case, the reconstructed map exhibits a lower integrated
flux in the asymmetric source region, the excess being redistributed as
a concentric annulus about the same radial distance.
By limiting the amount of azimuthal smearing\index{azimuthal smearing}
it is possible to alleviate this effect and to start recovering
azimuthal information in the accretion disc\index{accretion disc}.
Two prescriptions in this regard were proposed. Rutten et al. \cite{r93}
limited the amount of azimuthal smearing\index{azimuthal smearing}
by averaging over a polar Gaussian weight
function\index{weight function} of {\em constant angle} $\Delta_\theta$
along the map
\footnote{$R_j$ and $R_k$ are the distances from pixels $j$ and $k$ to the
  centre of the disc; $\theta_{jk}$ is the azimuthal angle between pixels $j$
  and $k$; $s_{jk}$ is the arc-length between pixels $j$ and $k$.},
\begin{equation}
\omega_{jk} = \exp \left[ - \frac{1}{2} \left( \frac{(R_j - R_k)^2}{\Delta_R^2}
  + \frac{\theta_{jk}^2} {\Delta_\theta^2} \right) \right]
\label{eq:angle}
\end{equation}
\noindent while Baptista et al.\ \cite{bsh} chose to use a polar Gaussian function of
{\em constant arc length} $\Delta_s$ through the map,
\begin{equation}
\omega_{jk} = \exp \left[ - \frac{1}{2} \left( \frac{(R_j - R_k)^2}{\Delta_R^2}
    + \frac{s_{jk}^2}{\Delta_s^2} \right) \right] \,\, .
\end{equation}
\noindent The reconstructions in Figs.\,\ref{fig:4}(b)-(d) used the
default function\index{default function} of Eq.\,\ref{eq:angle}. In particular,
the reconstructions in Figs.\,\ref{fig:4}(a) and \ref{fig:4}(b) were
obtained from the same input light curve\index{light curve}. This comparison
illustrates that the default of limited azimuthal
smearing\index{azimuthal smearing} provides a much better reconstruction
of the central brightness distribution while still allowing to recover
the location of a bright spot\index{bright spot} at disc
rim\index{disc rim}, although with some azimuthal smearing.
%
%%%%%%%%%% FIGURE 4 %%%%%%%%%%%%%
%
%%%%%%%%%%%%%%%%%%%%%%%%%%%%%%%%%

The concept of default map was extended with the introduction of a
double default function\index{default function} \cite{bob97,spruit94,b05},
\begin{equation}
D_j = ({D1_j})^n  \, ({D2_j})^m
\end{equation}
\noindent (where $D1_j$ and $D2_j$ are separate default
functions\index{default function} and $n+m=1$), together with the idea of
a negative default function (e.g., $m<0$). While a positive default
function\index{default function} steers the MEM\index{maximum entropy}
solution {\em towards} its intrinsic property (e.g., axi-symmetry), a
negative default function\index{default function} may be used to drive
the MEM\index{maximum entropy} solution {\em away} from its intrinsic
property (e.g., the presence of the undesired criss-crossed arcs)
\cite{spruit94}. The combination of a positive axi-symmetric default
function\index{default function} with a negative criss-crossed arcs default
function\index{default function} was key to allow the recovery of accretion
disc\index{accretion disc} spiral\index{spiral shock} structures with
eclipse mapping\index{eclipse map} \cite{harlaftis04,b05}. The eclipse
maps\index{eclipse map} shown in Figs.\,\ref{fig:demo}, \ref{fig:3} and
\ref{fig:5} were obtained with this double default setup.

The quality of an eclipse mapping\index{eclipse map} reconstruction is
tied to the quality of the input data light curve\index{light curve}.
Specifically, the ability to recover brightness sources and the spatial
resolution\index{spatial resolution} of an eclipse map\index{eclipse map}
depend on the phase (or, time) resolution and the signal-to-noise ratio
($S/N$) of the data light curve\index{light curve}. Baptista \& Steiner
\cite{bs91} provide an expression to compute the spatial
resolution\index{spatial resolution} of the eclipse map\index{eclipse map}
which matches the phase resolution of the data light curve\index{light curve}. 
Fig.\,\ref{fig:4} shows the effects of (i) degrading the phase
resolution\index{phase resolution} and (ii) lowering the $S/N$ of the
data light curve\index{light curve} on the quality of the resulting
eclipse map\index{eclipse map}. As a reference for comparison, the eclipse
map\index{eclipse map} in Fig.\,\ref{fig:4}(b) was obtained from a light
curve\index{light curve} of good phase resolution\index{phase resolution}
($Np=121$ data points) and high signal-to-noise ($S/N=100$).
Fig.\,\ref{fig:4}(c) shows that degrading the $S/N$ reduces the ability
to identify fainter brightness sources. Because of the reduced $S/N$,
the model light curve\index{light curve} is not forced to follow the
egress shoulder that signals the presence of the bright
spot\index{bright spot} at disc rim\index{disc rim}. As a consequence,
the eclipse map\index{eclipse map} has only a weak asymmetry at the
position of the bright spot. Fig.\,\ref{fig:4}(d) shows that the
effect of reducing the phase resolution\index{phase resolution} in
the light curve\index{light curve} is to reduce the spatial
resolution\index{spatial resolution} of the eclipse
map\index{eclipse map}, with an increase in the radial/azimuthal blur effects
(i.e., point sources would look increasingly out of focus with decreasing
phase resolution\index{phase resolution}).
The effects of low $S/N$ seem more dramatic than those of low phase
resolution.

\section{Error Propagation Procedures}
\label{sec:4}

\begin{figure}[t]
%\sidecaption
\includegraphics[scale=.55,angle=270]{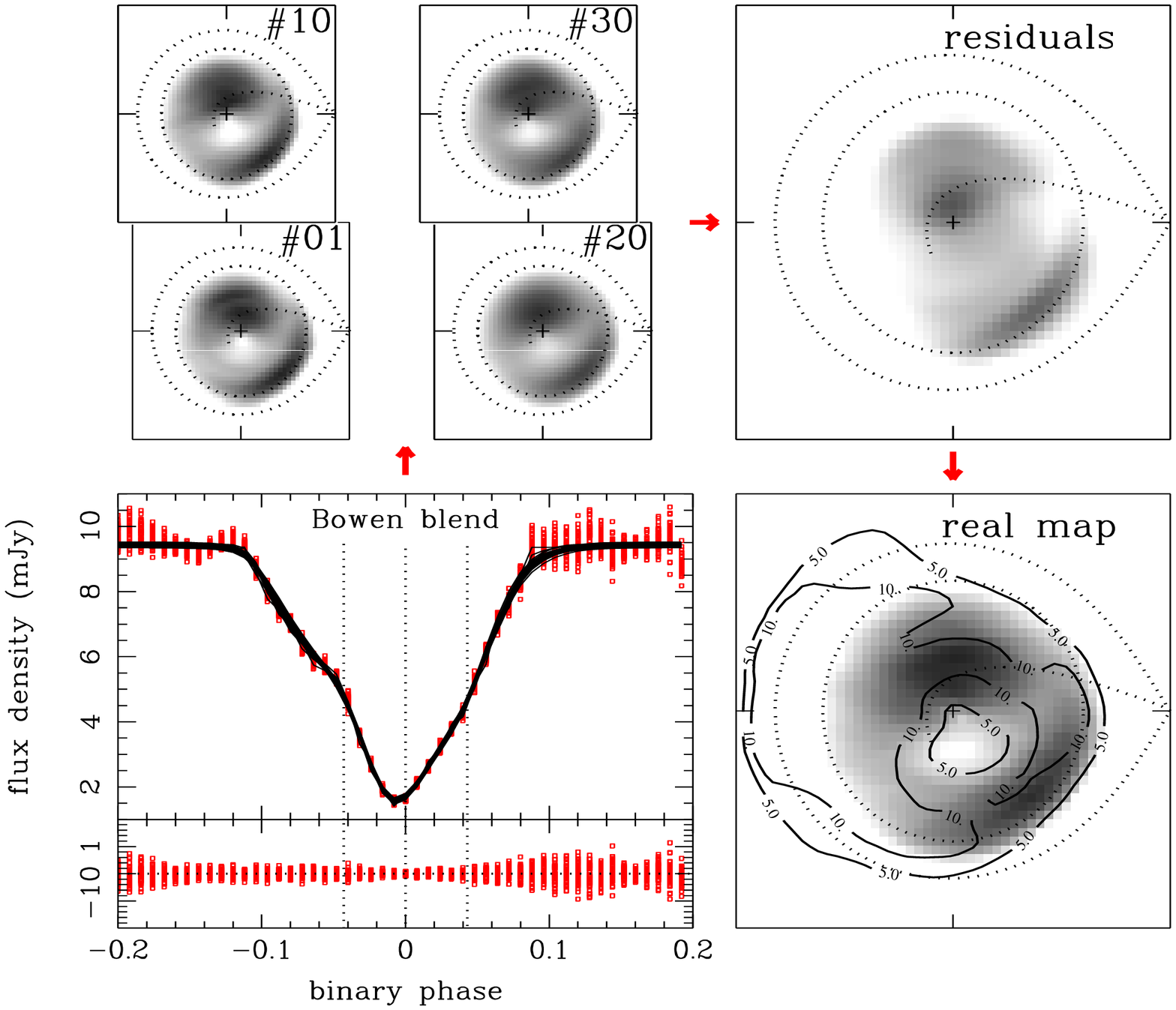}
\caption{Computing the statistical significance of an eclipse
  map\index{eclipse map} from the uncertainties in the data light
  curve\index{light curve}. Lower left panels: A set of 50 randomised
  light curves\index{light curve} (dots) and corresponding eclipse
  mapping\index{eclipse map} models (solid lines) superimposed in phase.
  The bottom panel shows these light curves after subtraction of the
  real data light curve\index{light curve}. The scatter depicts the
  uncertainty in the flux at each phase. Top left panels: a sample of
  4 of the eclipse maps\index{eclipse map} from the randomised light
  curves\index{light curve}, in a logarithmic grayscale.
  The notation is similar to that of Fig.\,\ref{fig:4}. Top right panel:
  the map of the standard deviations with respect to the mean intensity
  (the ``residuals'' map). Lower right panel: the eclipse
  map\index{eclipse map} from the real data light
  curve\index{light curve}. Contour lines for $S/N=5$ and 10 are
  overplotted; features in the eclipse maps\index{eclipse map} are
  statistically significant at or above the $5\,\sigma$ confidence level.}
\label{fig:5}
\end{figure}

Because the maximum-entropy\index{maximum entropy} eclipse
mapping\index{eclipse map} is a non-linear inversion method, it is
not possible to compute the uncertainties in the eclipse map directly
from the uncertainties in the data light curve\index{light curve}. Hence,
statistical uncertainties of an eclipse map\index{eclipse map} are
usually computed with a Monte Carlo error
propagation\index{error propagation} procedure \cite{rpt92}.
This is illustrated in Fig.\,\ref{fig:5}.
%
%%%%%%%%%% FIGURE 5 %%%%%%%%%%%%%
%
A set of N ($\geq 20$) artificial light curves is created from the
original data in which the flux at each phase is varied around the true
value according to a Gaussian distribution with standard deviation equals
to the uncertainty at that point. The artificial light
curves\index{light curve} are fitted with the eclipse
mapping\index{eclipse map} code to generate a set of randomised eclipse
maps\index{eclipse map}.
These are then combined to produce a ``residuals'' map by
taking the pixel-to-pixel standard deviation with respect to the mean
intensity. This yields the statistical uncertainty at each pixel.
For $N=20$, the uncertainty of the standard deviation is $<20\%$,
sufficient to illustrate the confidence limits of the eclipse
map\index{eclipse map},
whereas $N\geq 200$ is needed in order to bring the uncertainty in
the standard deviation down to $7\%$ \cite{harlaftis04}.
Uncertainties obtained with this procedure may be used to estimate
the errors in the derived radial brightness temperature\index{temperature}
and intensity distributions.
A map of the statistical significance (or the inverse of the
relative error) is obtained by dividing the true eclipse
map\index{eclipse map} by the ``residuals'' map
\cite{harlaftis04,bb04,b05}.

An alternative error propagation\index{error propagation} procedure
involves the use of simulations with the Bootstrap
technique\index{Bootstrap method} (e.g., \cite{bb08}).

\section{Applications}
\label{sec:5}

This section focuses on selected applications of the eclipse
mapping\index{eclipse map} method which illustrates some of the
possible scientific problems that have been and may be addresses
with it. The reader is referred to \cite{hm86,h93,w94,bap01} for
more comprehensive reviews of the results obtained with this technique.

\subsection{Spectral Mapping: Spatially-resolved Disc Spectra} %
\label{spec-map}\index{spatially resolved}

The eclipse mapping\index{eclipse map} method is capable of delivering
spatially-resolved spectra of accretion discs\index{accretion disc}
when the technique is applied to time-resolved eclipse spectrophotometric
data, providing the best example of its ability to disentangle the
light from different emission sources in the binary.
The time-series of spectra is divided up into numerous ($\sim 100$) spectral
bins and light curves\index{light curve} are extracted for each bin.
The light curves are then analysed to produce a series of monochromatic
eclipse maps\index{eclipse map} covering the 
whole spectrum.  Finally, the maps are combined to obtain the spectrum 
for any region of interest on the disc surface \cite{r93}.

\begin{figure}[t]
%\sidecaption
\scalebox{.45}{%
  \includegraphics[bb=1.3cm 6.2cm 20cm 24.5cm,clip,angle=270,scale=.55]
                  {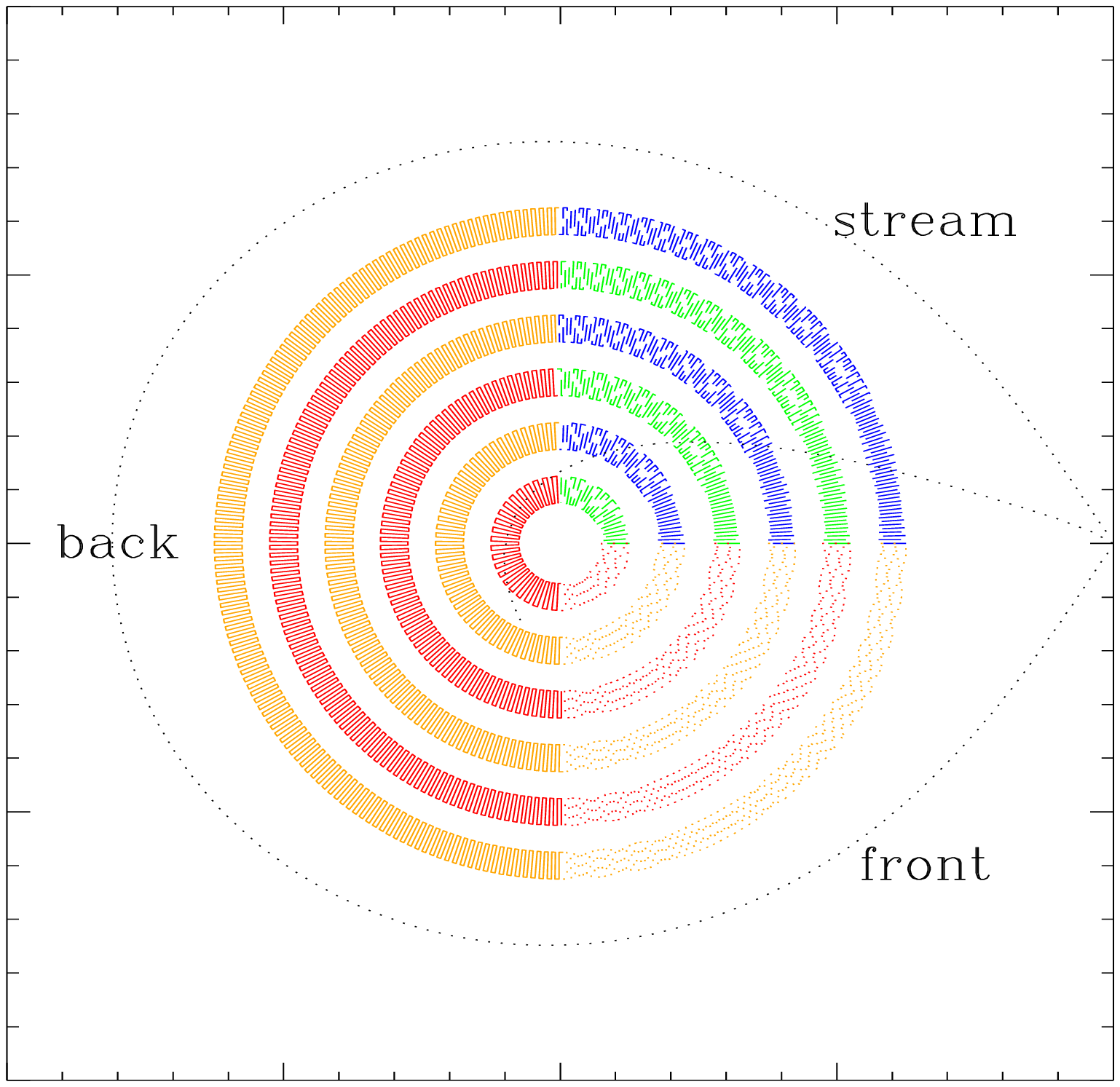} }%
\quad
\scalebox{.55}[0.6]{%
  \includegraphics[bb=2cm 4.5cm 19.5cm 20cm,angle=270,scale=.6]
                  {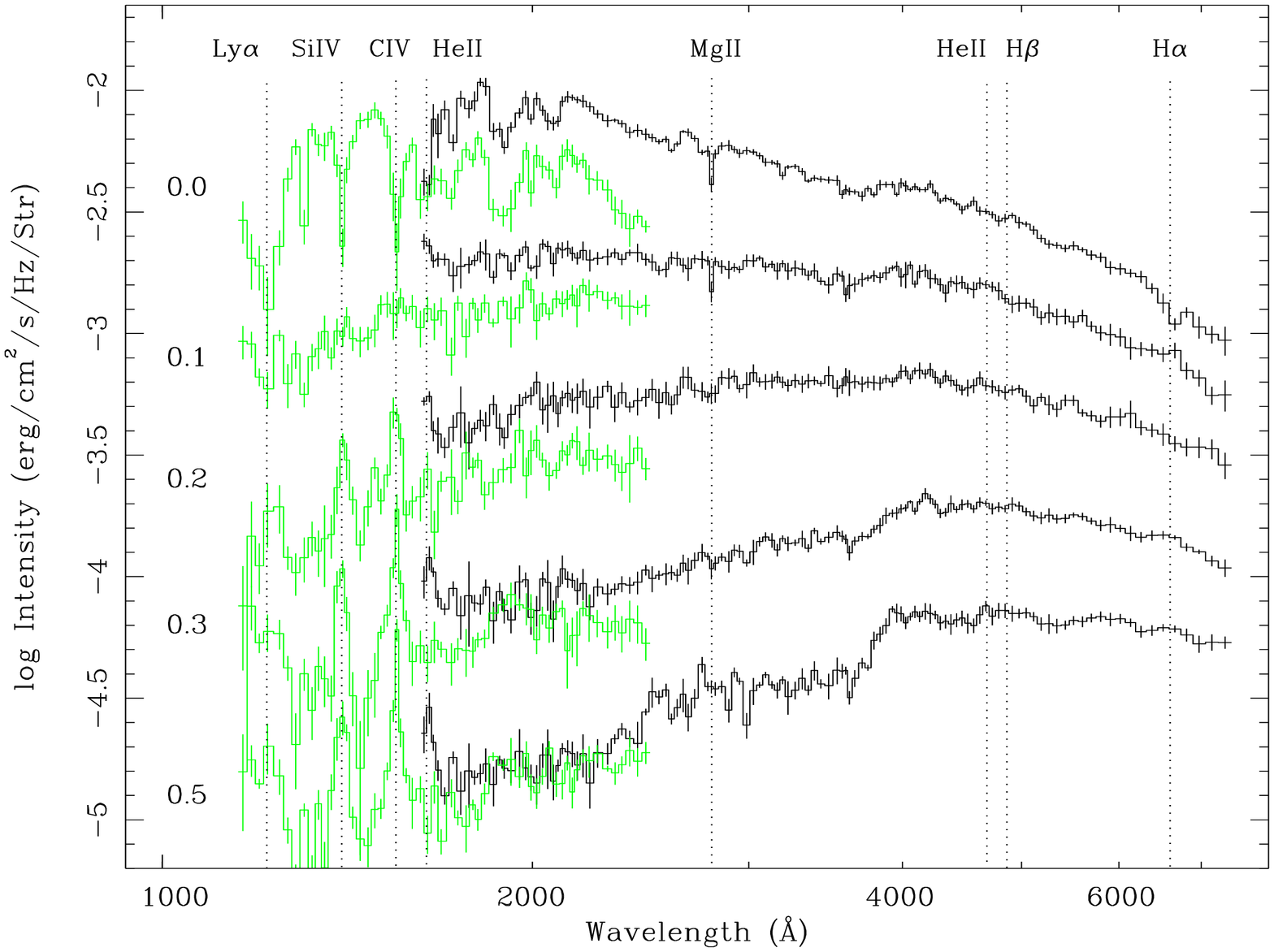} }
\caption{An example of spectral mapping, from HST/FOS UV-optical
  eclipse spectroscopy of the nova-like\index{nova-like} variable
  UX\,UMa\index{UX UMa}. Left panel: a schematic diagram of the eclipse
  map\index{eclipse map}, with the annular regions used to extract
  spatially resolved spectra\index{spatially resolved}. The disc is
  divided into three major azimuthal regions (the back side, the
  front side, and the quarter section containing the gas
  stream\index{accretion stream} trajectory), and into a set of 6 concentric
  annuli with radius increasing in steps of $0.1\; R_{L1}$ and of
  width\index{Lagrangian point} $0.05\; R_{L1}$.  Right panel:
  Spatially resolved spectra of the back region of the
  UX\,UMa\index{UX UMa} accretion disc\index{accretion disc} in two
  occasions (August 1994
  in green; November 1994 in black). The spectra were computed for the set
  of concentric annular sections shown in the left panel.
  The most prominent line transitions are indicated by vertical dotted
  lines. From \cite{b98}. }
\label{fig:6}
\end{figure}

The spectral mapping analysis of nova-like\index{nova-like} variables 
\cite{r93,r94,b98,bssh,gthesis} shows that their inner accretion
disc\index{accretion disc} is characterised by a blue
continuum\index{continuum emission} filled with absorption bands and lines
which cross over to emission with increasing disc radius (Fig.~\ref{fig:6}).
The continuum\index{continuum emission} emission becomes progressively
fainter and redder as one moves outwards, reflecting the radial
temperature\index{temperature} gradient.
These high mass-accretion\index{accretion} ($\sim 10^{-8} - 10^{-9} \;
M_\odot$\,yr$^{-1}$) discs seem hot and optically
thick\index{optically thick} in their inner regions and cool and optically
thin\index{optically thin} in their outer parts.
%
%%%%%%%%%% FIGURE 6 %%%%%%%%%%%%%
%
The spectrum of the infalling gas stream\index{accretion stream} in
UX~UMa\index{UX UMa} \cite{b98} and UU~Aqr\index{UU Aqr} \cite{bssh} is
noticeably different from the disc spectrum at the same
radius, suggesting the existence of gas stream ``disc-skimming'' overflow
that can be seen down to $R\simeq (0.1-0.2)\; R_{L1}$\index{Lagrangian point}.

In contrast, spatially-resolved\index{spatially resolved} spectra of
low-mass accretion ($\sim 10^{-9} - 10^{-11}\, M_\odot$ yr$^{-1}$), quiescent
dwarf nova\index{dwarf nova}
show that the lines are in emission at all disc radii \cite{saito06}.
The observed differences between the spectra of the disc hemisphere farther
from and closer to the L1 point\index{Lagrangian point} might be
interpreted in terms of chromospheric emission\index{chromospheric emission}
from a disc with non-negligible opening angle\index{opening angle}
(i.e., a limb-brightening effect).

The spectrum of the uneclipsed component both for the
nova-like\index{nova-like} systems and quiescent dwarf novae shows strong
emission lines and the Balmer jump\index{dwarf nova}\index{Balmer jump}
in emission indicating that the uneclipsed light has an important
contribution from optically thin\index{optically thin} gas (e.g.,
\cite{bssh}). The lines and optically thin
continuum\index{continuum emission} emission are most probably emitted
in a vertically extended disc
chromosphere\index{chromosphere} + wind\index{wind} \cite{b95,knigge97}.
This additional source of radiation may be responsible both for flattening
the ultraviolet spectral slope and for filling in the Balmer jump
of the\index{Balmer jump} optically thick\index{optically thick} disc
spectrum, and might explain the historical difficulties
in fitting integrated disc spectrum with stellar atmosphere disc models
\cite{wade1,dous2,long2,knigge98}.

The uneclipsed spectrum of UX~UMa\index{UX UMa} at long wavelengths
is dominated by a
late-type spectrum that matches the expected contribution from the
mass-donor\index{donor star} star \cite{r94}. Thus, the uneclipsed
component provides an interesting way of assessing the spectrum of
the mass-donor\index{donor star} star in eclipsing
CVs\index{cataclysmic variable}. This is a line of research yet to be
properly explored, which could largely benefit from near- and
mid-infrared\index{infrared} spectroscopy of eclipses. \index{eclipse} 

Spectral mapping is also useful to isolate the spectrum of the WD,
represented by the central pixel of the eclipse map\index{eclipse map}.
A stellar atmosphere model fit to the extracted white
dwarf\index{white dwarf} spectrum for the
dwarf nova\index{dwarf nova} V2051\,Oph\index{V2051 Oph} yields the
WD temperature\index{temperature} and an independent
estimate of the distance to the binary \cite{saito06}.

\subsection{Time-lapse Mapping: Dwarf Nova Outbursts}
\label{time-map}\index{time-lapse mapping}

Eclipse maps\index{eclipse map} give snapshots of the accretion
disc\index{accretion disc} at a given time. 
Time-resolved eclipse mapping\index{eclipse map} may be used to track
changes in the disc structure, e.g., to assess variations in mass
accretion rate\index{accretion rate} or to follow the evolution of
the surface brightness\index{surface brightness} distribution through a 
dwarf nova (DN) outburst cycle\index{dwarf nova outburst}.

DN outbursts\index{dwarf nova outburst} are explained in terms of
either (i) the time dependent response of a viscous accretion
disc\index{accretion disc} to a burst of mass transfer from the
donor\index{donor star} star (the mass transfer instability
model\index{mass transfer instability model},
MTIM\index{mass transfer instability model}, e.g.,
\cite{bath}), or (ii) a limit-cycle behaviour driven by a
thermal-viscous disc-instability (the disc-instability
model\index{disc instability model}, DIM\index{disc instability model},
e.g., \cite{lasota,c93}), in which matter progressively accumulates in
a low viscosity disc (quiescence) until a critical surface
density\index{surface density} is reached at a given radius, causing
a heating wave\index{heating wave} to switch the disc to a high
viscosity\index{disc viscosity} regime (outburst) that allows the gas
to diffuse rapidly inwards and onto the white dwarf\index{white dwarf}.
There is a set of distinct predictions from these models that might be
critically tested with time-lapse eclipse\index{time-lapse mapping}
mapping \cite{b12}.

\begin{figure}
%\sidecaption
\includegraphics[scale=.57]{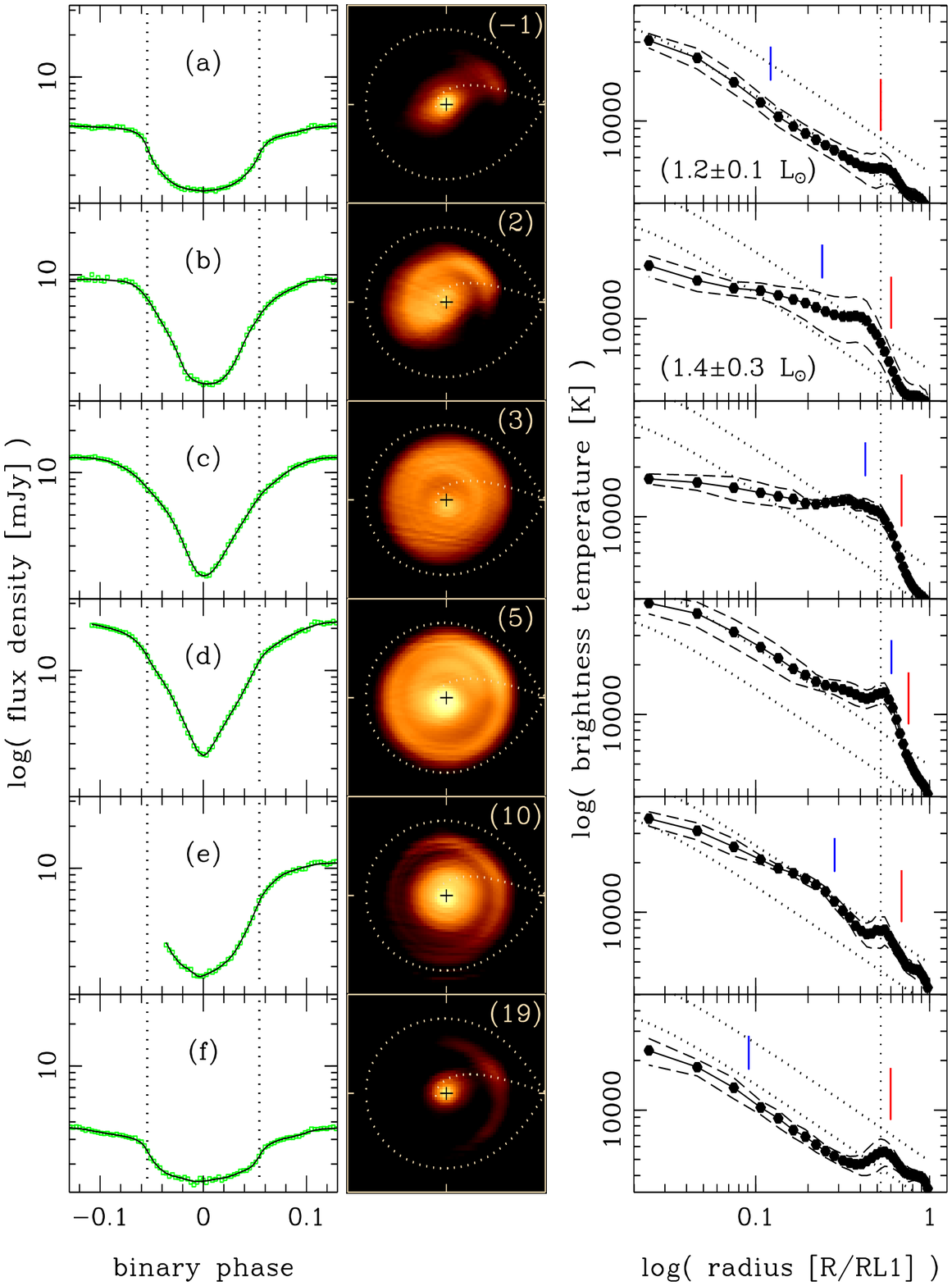}
\caption{Time-lapse\index{time-lapse mapping} eclipse
  mapping\index{eclipse map} of the dwarf nova EX~Dra\index{EX Dra} along
  its outburst cycle\index{dwarf nova outburst}. Left-hand panels:
  Data (dots) and model (solid line)
  light curves\index{light curve} in (a) quiescent, (b) early rise,
  (c) late rise, (d) outburst maximum\index{dwarf nova outburst}, (e)
  early decline, and (f) late decline stages.
  Vertical dotted lines mark ingress/egress phases of disc centre.
  Middle panels: eclipse maps\index{eclipse map} in a false colour blackbody
  logarithmic grayscale. The notation is similar to that of Fig.\,\ref{fig:demo}.
  The numbers in parenthesis indicate the time (in days) elapsed since
  outburst onset\index{dwarf nova outburst}. Right-hand panels:
  azimuthal-averaged radial brightness
  temperature\index{temperature} distributions for the eclipse
  maps\index{eclipse map} in the middle panels.
  Dashed lines show the 1-$\sigma$ limit on the average
  temperature\index{temperature} for a given radius.
  A dotted vertical line depicts the radial position of the bright
  spot\index{bright spot} in quiescence; vertical ticks mark the position
  of the outer edge of the disc (red) and the radial position at which the
  disc temperature\index{temperature} falls below 11000\,K (blue).
  Steady-state disc models for mass accretion rates\index{accretion rate}
  of $\log$ \.{M}$= -8.0$ and $-9.0 \;M_\odot\;$yr$^{-1}$ are plotted as
  dotted lines for comparison. Numbers in
  parenthesis list the integrated disc luminosity. From \cite{bc01}.}
\label{fig:7}
\end{figure}

An example of time-lapse\index{time-lapse mapping} eclipse
mapping\index{eclipse map} is given in Fig.\,\ref{fig:7}. Eclipse
maps\index{eclipse map} covering the full outburst
cycle\index{dwarf nova outburst} of the dwarf nova\index{dwarf nova}
EX~Dra\index{EX Dra} \cite{bc01} show the formation of a one-armed
spiral\index{spiral shock} structure at the early stages of the
outburst\index{dwarf nova outburst} and reveal how the disc expands
during the rise until it fills most of the primary Roche
lobe\index{Roche lobe} at maximum light.
During the decline stage, the disc becomes progressively fainter until
only a small bright region around the WD is left at minimum light.
%
%%%%%%%%%% FIGURE 7 %%%%%%%%%%%%%
%
The evolution of the radial temperature\index{temperature} distribution
shows the presence of an outward-moving heating wave\index{heating wave}
during rise and of an inward-moving cooling wave in the decline. 

The temperatures\index{temperature} of the EX\,Dra\index{EX Dra}
outbursting\index{dwarf nova outburst} disc are above those expected for
the hot, ionised outbursting disc gas in the
DIM\index{disc instability model} framework \cite{bible,lasota}, and
the cooling wave decelerates as it travels inwards \cite{bc01}. Since
these results are consistent with both DIM\index{disc instability model}
and MTIM\index{mass transfer instability model}, they offer no power to
discriminate between these models. However, the same eclipse mapping
study reveals that the radial temperature\index{temperature} distribution
in quiescence follows the $T \propto R^{-3/4}$ law of opaque steady-state
discs (Fig.\,\ref{fig:7}(a)), and that this quiescent state is reached
only 2\,d after the end of the outburst\index{dwarf nova outburst}.
Both results indicate that the viscosity\index{disc viscosity} of its
quiescent disc is as large as in outburst\index{dwarf nova outburst} -- in
contradiction with DIM\index{disc instability model} and in agreement
with predictions of the MTIM\index{mass transfer instability model}.
In addition, the early rise map shows evidence of enhanced gas
stream\index{accretion stream}
emission, indicative of an enhanced mass transfer rate at this early
outburst stage (Fig.\,\ref{fig:7}(b)). The integrated disc luminosity at
early rise ($1.4\pm 0.3\, L_\odot$) is comparable to that in quiescence
($1.2 \pm 0.1\, L_\odot$), and is not enough to support the idea that the
enhanced mass transfer could be triggered by an increased
irradiation\index{irradiation} of the mass donor\index{donor star} by the
accretion disc.\index{accretion disc} This led to the conclusion that
the observed enhanced mass transfer at early rise is not a consequence
of the ongoing outburst\index{dwarf nova outburst}, but its cause --
suggesting that the outbursts\index{dwarf nova outburst} of
EX~Dra\index{EX Dra} are powered by bursts of mass transfer \cite{b12}.

Time-lapse\index{time-lapse mapping} eclipse mapping\index{eclipse map}
studies were also performed for the dwarf novae\index{dwarf nova}
Z\,Cha\index{Z Cha} \cite{wo88}, OY\,Car\index{OY Car} \cite{r92a, bbb96},
IP\,Peg\index{IP Peg} \cite{bob97}, V2051\,Oph\index{V2051 Oph}
\cite{bsfb07}, and HT\,Cas\index{HT Cas} \cite{zach99,feline05}.

\subsection{Flickering Mapping\index{flickering map}: Revealing the Disc Viscosity} %
\label{flick-map}

Flickering\index{flickering} is the intrinsic brightness fluctuation seen
in light curves\index{light curve} of T\,Tau stars\index{T Tau star}
\cite{herbst99}, mass-exchanging binaries \cite{aug92,bruch00} and active
galactic nuclei\index{active galactic nucleus} \cite{garcia99}.
Optical studies suggest there are two different sources of
flickering\index{flickering} in CVs\index{cataclysmic variable}: (i) the
stream-disc impact region\index{bright spot} (possibly because of unsteady
mass inflow or post-shock turbulence \cite{wn1971,shu76}) and/or (ii)
turbulent inner disc regions (possibly because of
magneto-hydrodynamic\index{magneto-hydrodynamic} (MHD) turbulence,
\index{magneto-hydrodynamic} \index{turbulence} unsteady WD
accretion\index{accretion} or events of magnetic
reconnection\index{magnetic reconnection}
at the disc atmosphere \cite{ga92,bruch92,kawa00}). With its ability
to spatially-resolve\index{spatially resolved} and to disentangle
different sources, flickering mapping\index{flickering map} has been a
useful tool to confirm and extend this scenario. As an added bonus,
if the disc-related flickering\index{flickering} is caused
by MHD turbulence, \index{turbulence}\index{magneto-hydrodynamic}
one may infer the radial run of the disc viscosity\index{disc viscosity}
parameter $\alpha_\mathrm{ss}$ \cite{ss73} from the relative
flickering\index{flickering} amplitude, $(\sigma_\mathrm{D}/D)$ \cite{ga92},
\begin{equation}
  \alpha_{ss}(R) \simeq 0.23\, \left[ \frac{R}{50\,H} \right] \left[
  \frac{\sigma_\mathrm{D}(R)} {0.05\,D(R)} \right]^2 \, ,
\label{flick-ampl}
\end{equation}
where $H$ is the disc scale height.

From a large, uniform ensemble of light curves\index{light curve} of a
given CV\index{cataclysmic variable} it is
possible to separate the steady-light component (the average flux in a
given phase bin), low- and high-frequency flickering\index{flickering}
amplitudes (the scatter with respect to the average flux, computed with
the ensemble \cite{bennie96} and single \cite{bruch92} methods) as a
function of binary phase, to derive corresponding maps of surface
brightness\index{surface brightness}
distributions from their eclipse shapes and, thereafter, to compute the
radial run of the relative amplitude of the disc
flickering\index{flickering} component \cite{bb04,bb08}.

Flickering mapping\index{flickering map} of the DN\index{dwarf nova}
V2051\,Oph\index{V2051 Oph} reveals that the low-frequency
flickering\index{flickering} arises mainly in an overflowing gas
stream\index{accretion stream} and is associated with the mass transfer
process. The high-frequency flickering\index{flickering} originates
in the accretion disc\index{accretion disc} and has a relative amplitude
of a few percent, independent of disc radius and brightness level,
leading to large values $\alpha_\mathrm{ss}\simeq (0.1-0.2)$ at all
disc radii \cite{bb04}.

\begin{figure}[t]
%\sidecaption
\includegraphics[bb=1cm 0.5cm 19.5cm 16cm,scale=.48,angle=270]
                  {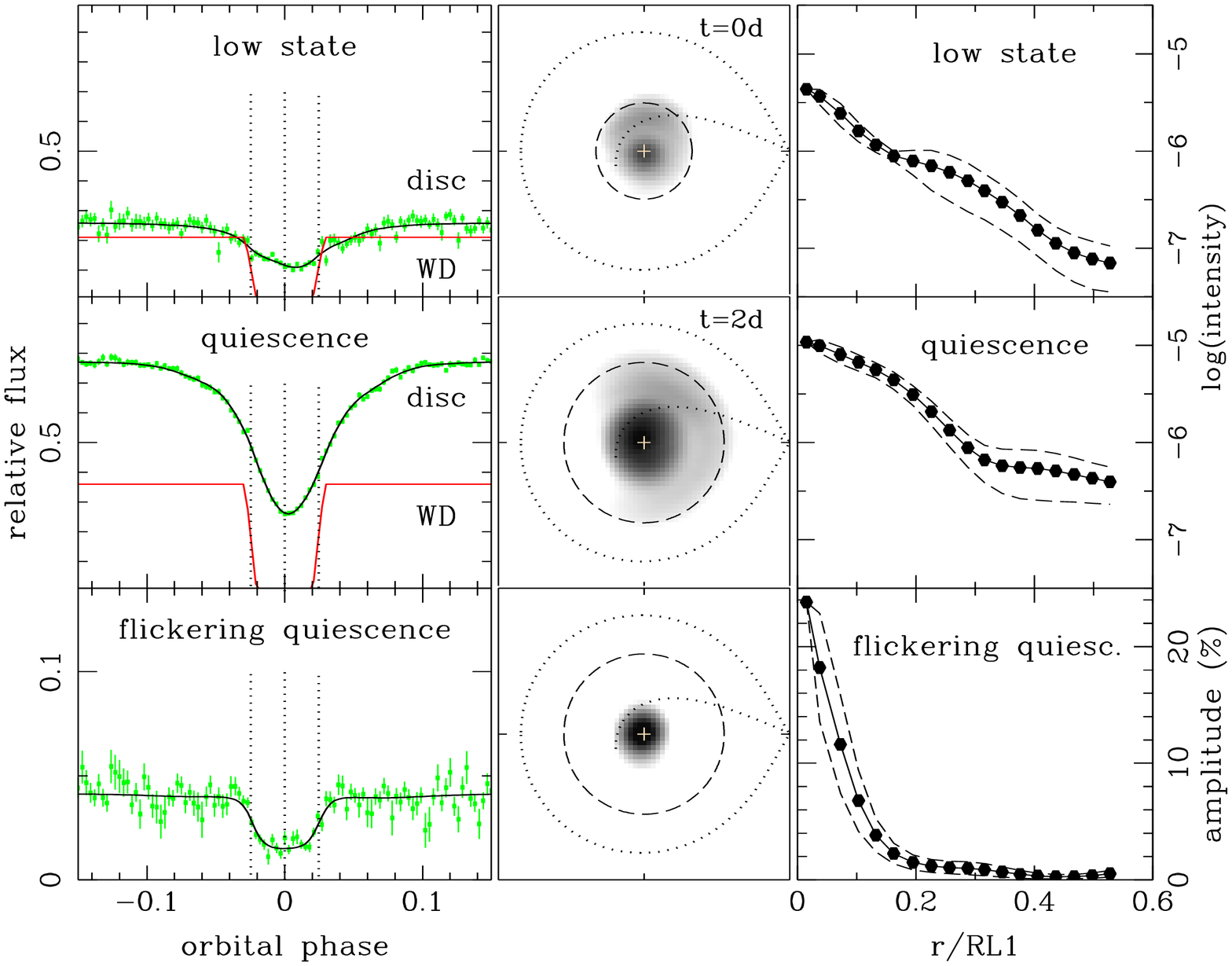}
\caption{ Left-hand panels: data (green dots) and eclipse
  mapping\index{eclipse map} model (black lines) disc light
  curves\index{light curve}, and the extracted light
  curve\index{light curve} of the WD (red lines). Top and middle
  panels show, respectively, the low state and quiescence steady-light
  component, while the lower panel shows the high-frequency
  ($f>3.3\,mHz$) flickering\index{flickering} curve. Middle panels: the
  corresponding eclipse maps\index{eclipse map} in a logarithmic grayscale.
  The notations is similar to that of Fig.\,\ref{fig:geom}.
  Dashed circles mark the outer disc radius in each case. Right-hand
  panels: azimuthal-averaged radial intensity (top, middle) and
  flickering\index{flickering} amplitude in percent (bottom)
  distributions for the eclipse maps\index{eclipse map} in the middle
  panels. Dashed lines show the 1-$\sigma$ limit on the average value
  for a given radius.}
\label{fig:8}
\end{figure}

Figure\,\ref{fig:8} shows the results of the eclipse
mapping\index{eclipse map} analysis of an extensive data set of optical
light curves\index{light curve} of the dwarf nova\index{dwarf nova}
HT\,Cas\index{HT Cas} \cite{b12}. These observations frame a 2\,d
transition from a low state (largely reduced mass transfer rate) back
to quiescence, allowing the application of both
time-lapse\index{time-lapse mapping} and flickering
mapping\index{flickering map} techniques to derive independent
estimates of $\alpha_\mathrm{ss}$.
In the low state, the gas stream\index{accretion stream} hits the disc
at the circularisation radius\index{circularisation radius} $R_\mathrm{circ}$
\cite{bible}, and the accretion disc\index{accretion disc} has its
smallest possible size. The disc fast viscous response to the onset of
mass transfer, increasing its brightness and expanding its outer radius
at a speed $v \simeq +0.4\,km\,s^{-1}$, implies $\alpha_\mathrm{ss}\simeq
0.3-0.5$. The newly added disc gas reaches the WD at disc centre soon
after mass transfer recovery ($\sim 1$\,d), also implying a large disc
viscosity\index{disc viscosity} parameter, $\alpha_\mathrm{ss}\simeq 0.5$.
Flickering mapping\index{flickering map} reveal a minor, low-frequency
BS-stream\index{accretion stream}\index{bright spot}
flickering\index{flickering} component in the outer disc, plus a main
disc flickering\index{flickering} component the
amplitude of which rises sharply towards disc centre (Fig.\,\ref{fig:8}),
leading to a radial dependency $\alpha_\mathrm{ss}(R) \propto R^{-2}$ with
$\alpha_\mathrm{ss}>0.1$ for $R<R_\mathrm{circ}$ -- in agreement with the
time-lapse\index{time-lapse mapping} results.
%
%%%%%%%%%% FIGURE 8 %%%%%%%%%%%%%
%

A similar analysis was performed for the DN\index{dwarf nova}
V4140\,Sgr\index{V4140 Sgr} \cite{b15}.
Eclipse mapping\index{eclipse map} in quiescence indicate that the
steady-light is dominated by emission from an extended disc with
negligible contribution from the WD, suggesting that efficient
accretion\index{accretion} through a high-viscosity\index{disc viscosity}
disc is taking place. 
Flickering maps\index{flickering map} show an asymmetric source at disc
rim\index{disc rim} (BS-stream\index{accretion stream}\index{bright spot}
flickering\index{flickering}) and an extended central source (disc
flickering\index{flickering}) several times larger in radius than the WD.
Unless the thin disc approximation breaks down, the relative amplitude
of the disc flickering\index{flickering} leads to
large $\alpha_\mathrm{ss}$'s in the inner disc regions ($\simeq 0.2-1.0$),
which decrease with increasing radius.

In contrast, in the nova-like\index{nova-like} UU\,Aqr\index{UU Aqr}
optical flickering\index{flickering} arises mainly in tidally-induced
spiral\index{spiral shock} shocks in its outer disc \cite{bb08}.
Assuming that the turbulent \index{turbulence} 
disc model applies, its disc viscosity\index{disc viscosity}
parameter increases outwards and reaches $\alpha_{ss}\sim 0.5$ at the
position of the shocks, suggesting that they might be an effective source
of angular momentum removal of disc gas.

\subsection{3D Eclipse Mapping: Disc Opening angle \& Superhumps} %
\label{3d-map}\index{eclipse map, 3D}

Standard eclipse mapping\index{eclipse map} assumes a simple flat,
geometrically thin disc model. Real discs may however violate this
assumption in the limit of high \.{M}. Disc half-opening
angles\index{opening angle} of $\beta \geq 4^o$ are
predicted for \.{M}$\geq 5 \times 10^{-9}\;M_\odot\,yr^{-1}$ \cite{mm82}.
At large inclinations\index{inclination} ($i\geq 80^o$) this may lead to
artificial front-back asymmetries in the eclipse map\index{eclipse map}
because of the different effective areas of surface elements in the near
and far sides of a flared disc as seen by an observer on Earth.
Furthermore, the assumption that the emitted radiation is independent
of orbital phase implies that any out-of-eclipse brightness change
(e.g., orbital modulation due to BS\index{bright spot} anisotropic
emission) has to be removed before the light curve\index{light curve}
can be analysed
\footnote{This is usually done by fitting a spline function to the
  phases outside eclipse, dividing the light curve\index{light curve} by
  the fitted spline, \index{eclipse} and scaling the result to the spline
  function value at phase zero (e.g., \cite{b95}).}.

A step to overcome these limitations is to go beyond the standard
assumptions allowing the eclipse mapping\index{eclipse map} surface to
become three-dimensional. This leads to 3D eclipse
mapping\index{eclipse map, 3D} \cite{r98}.
For example, with the inclusion of a disc rim\index{disc rim} in the
eclipse mapping\index{eclipse map} method, the out-of-eclipse modulation
can be modelled as the fore-shortening of an azimuthal-dependent brightness
distribution in the disc rim\index{disc rim} \cite{bob97}.
This procedure allows one to recover the azimuthal (phase)
dependency of the BS\index{bright spot} emission.
Moreover, motivated by the front-back asymmetry that appeared in the
flat-disc map and by the difficulties in removing the asymmetry with the
assumption of an uneclipsed component, \cite{rob99} introduced a flared
disc in their eclipse mapping\index{eclipse map} of ultraviolet light
curves\index{light curve} of Z\,Cha\index{Z Cha} at
outburst\index{dwarf nova outburst}. They found that the asymmetry
vanishes and the disc is mostly axi-symmetric for a disc half-opening
angle\index{opening angle} of $\beta= 6^o$.

\begin{figure}[t]
%\sidecaption
\includegraphics[bb=0cm 1.2cm 19.5cm 16cm,scale=.5,angle=270]{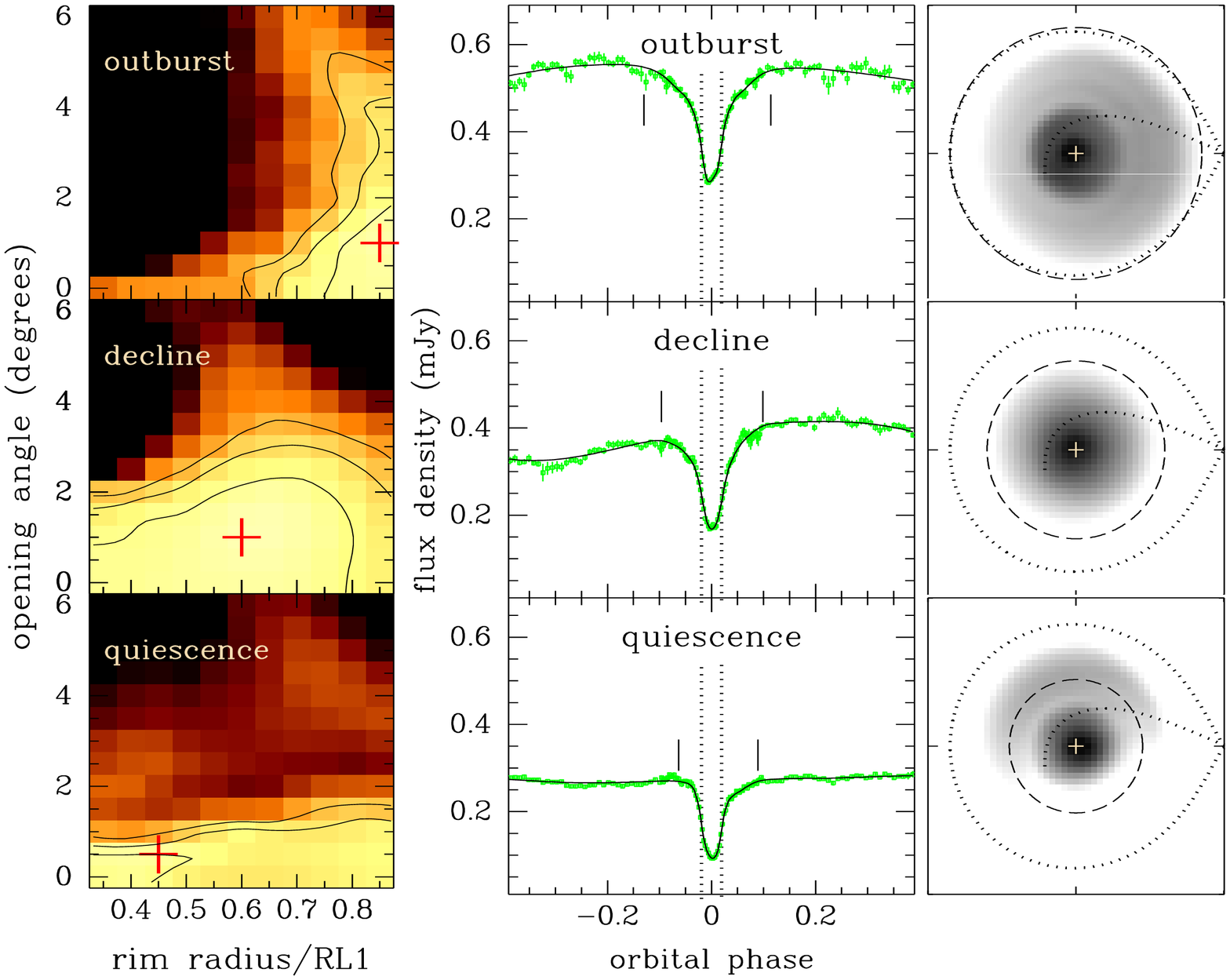}
\caption{ 3D eclipse mapping\index{eclipse map, 3D} of V4140\,Sgr. Left
  panels: entropy landscape for the outburst\index{dwarf nova outburst}
  (top), decline (middle) and quiescence (bottom) light
  curves in a blackbody false-colour scale. Regions in yellow have higher
  entropy\index{entropy}. A red cross marks the combination of disc
  half-opening angle\index{opening angle} $\beta$ and rim\index{disc rim}
  radius $R_d$ of highest entropy\index{maximum entropy}.  Middle panels:
  data (dots) and model (solid line) light curves\index{light curve} for
  the pair of $(\beta, R_d)$ values of highest entropy\index{maximum entropy}
  in each case. Vertical dotted
  lines mark the ingress/egress phases of the WD, while vertical tick marks
  depict the phases of eclipse ingress/egress. Right panels: corresponding
  eclipse maps\index{eclipse map} in a logarithmic grayscale. The notation
  is similar to that of Fig.\,\ref{fig:demo}. Dashed circles show the
  best-fit disc rim\index{disc rim} radius. From \cite{b15}.}
\label{fig:9}
\end{figure}

In 3D eclipse mapping\index{eclipse map, 3D}, the mapping surface usually
consists of a grid of $N1$ pixels on a conical surface centred at the WD
position and inclined at a half-opening angle\index{opening angle} $\beta$
with respect to the orbital plane\index{orbital plane}, plus
a circular rim\index{disc rim} of $N2$ pixels orthogonal to the orbital
plane\index{orbital plane} at a distance $R_d$ ($< R_\mathrm{L1}$)
from disc centre.  An \emph{entropy landscape} technique\index{entropy}
\cite{rd94} may be used to cope with
the extra degree of freedom that comes along with the additional geometry
parameters $\beta$ and $R_d$. Simulations \cite{b15} show that if an
eclipse mapping\index{eclipse map} reconstruction is performed with the
wrong choice of $\beta$ and $R_d$, the code develops artefacts in the
brightness distribution in order to compensate for the incorrect parameters.
Eclipse maps with these spurious, additional structures have lower
entropy\index{entropy} than the map with the correct choice of $\beta$ and
$R_d$. Therefore, the best-fit geometry can be found by searching the space
of parameters for the pair of ($\beta,R_d$) values of highest
entropy\index{maximum entropy} (Fig.\,\ref{fig:9}, left-hand panels).
%
%%%%%%%%%% FIGURE 9 %%%%%%%%%%%%%
%

An example of 3D eclipse mapping\index{eclipse map, 3D} is shown in
Fig.\,\ref{fig:9}, where the technique was applied to study the evolution
of the accretion disc\index{accretion disc} surface brightness of the
DN\index{dwarf nova} V4140\,Sgr\index{V4140 Sgr} in a
superoutburst\index{superoutburst} \cite{b15}.
The entropy landscape\index{entropy} analysis indicates that the accretion
disc\index{accretion disc} is geometrically thin both in outburst
($\beta= 1.0^o\pm 0.5^o$) and in quiescence ($\beta= 0.5^o\pm 0.5^o$);
it fills the primary Roche lobe\index{Roche lobe} in
outburst\index{dwarf nova outburst} and progressively shrinks to about
half this size in quiescence.
They also find that the disc is elliptical\index{elliptical disc} in
outburst\index{dwarf nova outburst} and decline, with an eccentricity
$e=0.13$. At both outburst\index{dwarf nova outburst} stages, the disc
orientation is such that superhump\index{superhump} maximum occurs when
the mass-donor\index{donor star} star is aligned with the bulge of the
elliptical disc\index{elliptical disc}. This lends observational support
for the tidal resonance instability model of superhumps\index{superhump}
\cite{whitehurst88,ho90,lubow94}.

Additional observational support for the presence of
elliptical\index{elliptical disc} precessing\index{precessing disc}
discs in CVs\index{cataclysmic variable} come from the eclipse
mapping\index{eclipse map} analyses of light curves\index{light curve}
of the DN\index{dwarf nova} Z\,Cha\index{Z Cha} in
superoutburst\index{superoutburst} \cite{o90} and of the permanent
superhumper\index{superhump} V348\,Pup\index{V348 Pup} \cite{rolfe00}.

\section{Summary}

Eclipse mapping\index{eclipse map} is a unique, powerful technique to
investigate:
\begin{itemize}

\item Accretion disc spectra (allowing one to separate the disc
  atmosphere emission at different distances from disc centre, and to
  isolate the spectrum of the WD, BS\index{bright spot}, a possibly
  outflowing disc wind\index{wind}, and even the faint, red
  mass-donor\index{donor star} star);

\item Accretion disc structures (such as gas
  stream\index{accretion stream} overflow, tidally-induced spiral
  shocks\index{spiral shock}, elliptical\index{elliptical disc}
  precessing\index{precessing disc} discs, and magnetic accretion curtains);

\item Time evolution of accretion discs\index{accretion disc} (tracing
  the mass and angular momentum redistribution throughout dwarf nova
  outbursts\index{dwarf nova outburst}, or to
  follow brightness changes caused by lighthouse effects in discs with
  fast spinning, magnetic WDs); \index{magnetic cataclysmic variable}

\item Accretion disc viscosity\index{disc viscosity} (either via
  flickering mapping\index{flickering map} or by measuring the velocity
  of transition waves during dwarf nova
  outbursts\index{dwarf nova outburst} with
  time-lapse\index{time-lapse mapping} mapping).
  
\end{itemize}

\begin{acknowledgement}
RB acknowledges financial support from CNPq/Brazil through grant 
no.\ 308\,946/2011-1.

\end{acknowledgement}

\end{document}